\documentclass[aps,prl,epsfigure,twocolumn,showpacs]{revtex4}
\usepackage{amsmath}
\usepackage{amstext}
\usepackage{latexsym}
\usepackage{psfig}
\usepackage{graphicx}
\usepackage{amsfonts}

\newcommand{\ket}[1]{\left\vert#1\right\rangle}
\newcommand{\modul}[1]{\left\vert#1\right\vert}

\newcommand{\one}{\mbox{$1 \hspace{-1.0mm}  {\bf l}$}}

\newcommand{\bra}[1]{\left\langle#1\right\vert}
\newcommand{\sand}[3]{\left\langle#1\vert#2\vert#3\right\rangle}

\newcommand{\valmed}[1]{\left\langle#1\right\rangle}

\begin{document}

\title{Complete conditions for entanglement transfer}
\author{M. Paternostro, W. Son and M. S. Kim}
\affiliation{School of Mathematics and Physics, The Queen's University,
Belfast BT7 1NN, United Kingdom}
\date{\today}

\begin{abstract}
We investigate the conditions to  entangle two qubits interacting
with local environments driven by a continuous-variable correlated
field. We find the conditions to transfer the entanglement from
the driving field to the qubits both in dynamical and steady-state
cases. We see how the quantum correlations initially present in
the driving field play a critical role in the
entanglement-transfer process. The system we treat is general
enough to be adapted to different physical setups.
\end{abstract}
%\pacs{03.67.Hk, 42.50.-p, 03.67.-a, 03.65.Bz}
\maketitle

%%%%%%%%%%%%%%%%%%%%%%%%%%%%%%%%%%%INTRODUZIONE%%%%%%%%%%%%%%%%%%%%%%%%%%%
%\section{Introduction}

Quantum networks of remote local processors, which are
interconnected by quantum and classical channels, have been
investigated to effectively perform quantum
computation~\cite{gottesman} and quantum communication~
\cite{duan}. A quantum repeater has also been proposed for
error-tolerant long-haul quantum communication~\cite{briegel}.
These schemes require a reliable channel to entangle remote nodes
in order to use in later steps of protocols. The usage of a light
field to implement a quantum channel is a natural choice because
of its handiness in generating and propagating entanglement
\cite{furusawa}.    However, we have witnessed that all optical
network is technically extremely challenging \cite{KLM}.  On the
other hand, static qubits such as the hyperfine structure of atoms
are easily accessible and manipulative by means of external
excitations.  Therefore, it may be an optimal strategy to use an
optical quantum channel to bring quantum correlation to two remote sites
of static qubits, where the entanglement is subsequently utilized
for quantum information processing. It is, thus, evident that the
study of entanglement transfer from an optical field of a
continuous-variable (CV) system to a static qubit system has a
primary importance.

Recently, such entanglement generation on the pair of remote
qubits has been studied through the indirect interaction via
projective measurement~\cite{plenio}, two-mode squeezed driving
field~\cite{sp,kraus} and a non-Markovian \cite{braun} and a
Markovian environments \cite{benatti}. Although these schemes
successfully demonstrate the situation for entanglement generation
on remote qubits, the complete physical requirements for the
possible creation of entanglement are unknown.  In this paper, we
investigate the sufficient and necessary conditions to induce
entanglement on two remote qubits, by means of their respective
linear interactions with a two-mode driving field.

From its definition, entanglement between any two systems cannot
be created by local unitary operations alone.  Thus, when the
driving field is separable, there is no way to generate the
entanglement between the two qubits. The natural least constraint
of the entanglement of the qubits is the entanglement of the state
for the quantum channel. However it is not clear if entanglement
of the driving field can always be transferred to the static
qubits. We study both the dynamical and the steady-state cases.

\begin{figure} [b]
\centerline{\psfig{figure=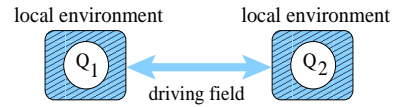,width=5.0cm,height=1.3cm}}
\caption{Scheme of the system considered.  Two individual qubits
($Q_{1},\,Q_{2}$), each interacting with their local environment,
are driven by a quantum correlated field.} \label{principio}
\end{figure}

%%%%%%%%%%%%%%%%%%%%%%%%%%%%%%%%%%%MASTEREQUATION%%%%%%%%%%%%%%%%%%%%%%%%%%%
{\it Master equation - } We analyze the dynamics of two remote
qubit systems sketched in Fig.~\ref{principio}.
Each qubit of its ground $\ket{g}_{i}$ and excited $\ket{e}_{i}$
states ($i=1,\,2$) interacts with its own local environment of a
single-mode bosonic system. We will refer to local environments by
modes $a$ and $b$. Static qubits such as ions, atoms or quantum
dot devices are isolated from uncontrollable real-world
environment. A coupling of the static qubit with the driving field
is thus assumed to be through their small local environments which
isolate the qubits from the uncontrollable environment. We model
each qubit-bosonic interaction by a resonant Hamiltonian
($\hbar=1$), $\hat{H}_{a1}+\hat{H}_{b2}$, where
$\hat{H}_{a1}=\Omega_{a1}\left
(\hat{\sigma}^{+}_{1}\hat{a}+\hat{a}^{\dag}\hat{\sigma}^{-}_{1}\right)$
(analogously for $\hat{H}_{b2}$). Here, $\hat{a}$ and
$\hat{a}^\dag$ are standard bosonic operators,
$\hat{\sigma}^{+}_{1}=(\hat{\sigma}^{-}_{1})^{\dag}=\ket{e}_{1}\!\bra{g}$
and $\Omega_{a1}$ is the Rabi frequency. The dynamics of the two
qubits is guided by an external broadband two-mode driving field
(bandwidth $\Delta\omega_{ext}$). The coupling between modes
$a,\,b$ and the external driving field is written as
$\hat{V}(t)=\sum_j\kappa_{j}\left(\hat{a}\hat{\cal
G}_{aj}^{\dagger}(t){e}^{i\omega_{a}t}+\hat{a}^{\dagger}\hat{\cal
G}_{aj}(t){e}^{-i\omega_{a}t}+a\leftrightarrow{b}\right)$, where
$\hat{\cal G}_{aj}(t)=\hat{a}_j\mbox{e}^{i\omega_j t}$ with
$\hat{a}_j$ the annihilation operator of the driving field in
frequency $\omega_j$, is coupled at rate $\kappa_{j}$ to mode $a$.
Here, $\omega_{a,b}$ are the frequencies of modes $a$ and $b$. In
the weak coupling limit
$\kappa_{j}\ll\omega_{a,b},\Delta\omega_{ext}$ ($\forall{j}$), we
use second-order perturbation theory and the first Born-Markov
approximation. The evolution of modes $a,\,b$ can, then, be
described by a Liouvillian super-operator ${\hat{\cal L}}$
involving only the {\em second-order moments of quadrature
variables} for the driving field ~\cite{wallsmilburn} in its
carrier frequency $\omega_0$, which is set to be resonant to the
qubit transition, {\em i.e.}, $\omega_0=\omega_a=\omega_b$. The
coupling rate in this frequency is denoted by $\kappa$. The
dynamics of the total system can, thus, be described by the master
equation
$\partial_{t}\rho=-i[\hat{H}_{a1}+\hat{H}_{b2},\rho]+{\cal
\hat{L}}\rho=\left(\hat{\cal L}_{o}+\hat{\cal L}\right)\rho$, with
$\rho$ the density matrix of the qubits+local environments system.

In order to characterize the master equation, we need to consider
only the second-order quadrature moments matrix for the driving
field which can be written by a matrix ${\bf M}$ .  The matrix
elements are $M_{\alpha\beta}=\langle\{\hat{x}_\alpha,
\hat{x}_\beta\}\rangle$ with quadrature operator vector $\hat{\bf
x}=(\hat q_1, \hat p_1, \hat q_2,\hat p_2)$. Without changing the
entanglement structure of the driving field, a very general real
matrix $\bf M$ can be transformed into the simple form by local
unitary operations~\cite{simon} (no matter the driving field being
Gaussian or non-Gaussian)
\begin{equation}
\label{variancematrix}
{\bf M}=
\begin{pmatrix}
{\bf n}&{\bf c}\\
{\bf c}&{\bf m}
\end{pmatrix}
\end{equation}
with ${\bf n}=n\one_{2}$, ${\bf m}=m\one_{2}$  ($n,m\geq{0}$) that describe the
local properties of each mode and ${\bf c}=diag[c_{1},\,c_{2}]$
that accounts for the inter-mode correlations.

We are interested in the qubit evolution so to eliminate the
modes $a,\,b$. This can be done using an adiabatic elimination
procedure valid in the {\it{weak-coupling regime}}
$\kappa\gg\Omega_{a1},\Omega_{b2}$. In this case, the dynamics of
the modes interacting with the driving field is much faster than
their interaction with the qubits. The qubits see modes $a$ and
$b$ in a steady state $\rho_{ss}$ not affected by the qubit-modes
dynamics. The adiabatic elimination proceeds by defining a
{\it{projection}} operator as ${\cal
P}\rho=\rho_{ss}\otimes{Tr}_{ab}(\rho)=\rho_{ss}\otimes\rho_{12}$.
Here $\rho_{12}$ is the density matrix of the qubits. Using the
property ${\cal P}\hat{\cal L}_{o}{\cal P}\rho=0$, the reduced
master equation takes the form
$\partial_{t}\rho_{12}={Tr}_{ab}\left\{\hat{\cal
L}_{o}\int^{\infty}_{0}e^{\hat{\cal L}t}\hat{\cal
L}_{o}\left(\rho_{ss}\otimes\rho_{12}\right)dt\right\}$. It is
straightforward to find that the dynamics of the qubits is fully
described by the effective Liouvillian
\begin{equation}
\label{benatti} \hat{\cal L}_{e}\rho_{12}=
\sum^{4}_{\alpha,\beta=1}D_{\alpha\beta}\left(\hat{{\cal
O}}_{\alpha}\rho_{12}\hat{{\cal
O}}_{\beta}-\frac{1}{2}\left\{\hat{{\cal O}}_{\beta}\hat{{\cal
O}}_{\alpha},\rho_{12}\right\}\right),
\end{equation}
with $\hat{{\cal O}}_{\alpha}= \sigma_{\alpha}\otimes\one$ for
$\alpha=1,2$ and $\hat{{\cal
O}}_{\alpha}=\one\otimes\sigma_{\alpha-2}$ for $\alpha=3,4$,
$\sigma_{1,2}$ the $x$ and $y$ Pauli operators.  The Kossakowski
matrix is ${\bf D}=\{\{{\bf A}, {\bf C}\},\{{\bf C}^\dag, {\bf
B}\}\}$, where ${\bf A}={\bf A}^\dag$, ${\bf B}={\bf B}^\dag$ and
${\bf C}$ are $2\times2$ matrices~\cite{benatti}. For the driving
field with its second-order moments as in
Eq.~(\ref{variancematrix}), we get
\begin{equation}
\label{benattimatrices}
{\bf A}=\frac{\gamma_{1}}{4}
\begin{pmatrix}
n&i\\
-i&n\\
\end{pmatrix}\,
{\bf B}=\frac{\gamma_{2}}{4}
\begin{pmatrix}
m&i\\
-i&m\\
\end{pmatrix}\,
{{\bf C}}=\frac{\sqrt{\gamma_{1}\gamma_{2}}}{4}
\begin{pmatrix}
c_{1}&0\\
0&c_{2}\\
\end{pmatrix}
.
\end{equation}
We have introduced the effective decay rate
$\gamma_{1,2}=2\Omega_{a1,b2}^2/\kappa$, resulting from the
adiabatic elimination. The map described by Eq.~(\ref{benatti}) is
completely positive (CP) iff ${\bf D}\geq{0}$. The interaction
model we are using does not contain phase-damping processes so
that in $\hat{\cal L}_{e}\rho_{12}$ the terms depending on the $z$
Pauli operator are absent.  Otherwise, the Kossakowski matrix we
have, describes a general Markovian interaction of two qubits with
their local environments. {\bf C} is a real matrix because of the
constraint of local interaction between qubits and their
respective environments.

It is possible to characterize the  entanglement capabilities of
the environment-mediated interaction of the qubits treated here.
In what follows we use the entanglement measure based on
negativity of partial transposition (NPT), defined by ${\cal
E}_{NPT}=-2\lambda^{-}_{i}$, where $\lambda^{-}_{i}$ is the
negative eigenvalue of the partially transposed density matrix
$\rho^{T_{2}}_{12}$ ($T_{2}$ indicates partial transposition with
respect to qubit $2$)~\cite{zyczkowski}. NPT is a necessary and
sufficient condition for entanglement of a bipartite qubit
system~\cite{Horodecki}.

According to ref.~\cite{benatti}, a  sufficient
condition to entangle the qubits, which follow the dynamics
described by Eq.~(\ref{benatti}), is
\begin{equation}
\left({{\bf u}^\dag}{{\bf A}}{\bf u}\right)\left({\bf v^\dag}{{\bf
B}^{T}}{\bf v}\right)<\modul{{\bf u^\dag}{\bf C}{\bf v}}^2.
\label{benatti-c}
\end{equation}
Here ${\bf u}=(\cos2\theta,-i)^T$ and ${\bf v}=(\cos2\varphi,i)^T$
carry information on the generic initial states of qubits, which
are unitarily rotated by the angles $\theta$ and $\varphi$ around
the $\bf z$ axes of their Bloch spheres. We note that while ${\bf
A,\,B}$ and ${\bf C}$ depend just on the form of the reduced
master equation, the condition to entangle the qubits depends on
their initial conditions via the vectors ${\bf u}$ and ${\bf v}$.
We use this condition now but will assess it later for the
steady-state entanglement condition.

So far the treatment has been very general. However, to analyze
the entanglement condition clearly, we restrict ourselves to the
case when $c_1=-c_2=c>0$ in Eq.~(\ref{variancematrix}) from now
on. In fact, this case covers most of the entangled CV states,
which have been studied, including {\em Gaussian} noisy two-mode
squeezed states and beam-splitted two single-mode squeezed states
and {\em non-Gaussian} entangled coherent states after local
unitary operations. In these conditions and for $\gamma_{1}\neq\gamma_{2}$,
the map in Eq.~(\ref{benatti}) is CP iff
$c^{2}\leq{\min}\{(m-1)(n+{1}),(m+1)(n-1)\}$. We find that, if the
qubits are initially in their ground states, the sufficient
condition (\ref{benatti-c}) for entanglement becomes
\begin{equation}
\label{mycondition} (n-1)(m-1)<c^2.
\end{equation}
If this condition is satisfied for the quantum channel,
entanglement is created between two remote qubits for some period
of time. The uncertainty principle for the quantum channel can be
written in the following compact form: ${\bf M} - {\boldsymbol
\sigma}_y\oplus {\boldsymbol \sigma}_y\geq 0$. For the case of a
Gaussian field, if and only if the partial transposition of their
density matrix violates the uncertainty principle, the field is
entangled \cite{simon} and this condition reduces to
Eq.~(\ref{mycondition}) \cite{leekimmunro}. Therefore, we find
that {\em two remote qubits can be entangled for some periods
during its linear interaction with local environments if and only
if the Gaussian quantum channel of $|c_1|=|c_2|$ is entangled} by
appropriately choosing initial states of the qubits~\footnote{For
$\modul{c_{1}}\neq\modul{c_{2}}$ and
$\rho_{12}(0)=\ket{gg}_{12}\!\bra{gg}$, there are values of
$n,m,c_{1},c_{2}$ for which, while the drive is entangled, the
qubits are not. Thus, a drive having $\modul{\valmed{{\hat
q}_{1}{\hat q}_{2}}}=\modul{\valmed{{\hat p}_{1}{\hat p}_{2}}}$
optimizes the entanglement transfer because, in this case,
whenever entanglement is in the drive, it can be transferred to
the qubits.}. The second-order moment matrix
(\ref{variancematrix}) is obtained from a general case by local
unitary operations of the field. This can be interpreted as
transforming the initial qubit states as leaving the quantum
channel in the simple form (\ref{variancematrix}). For the
dynamic entanglement of the qubits, we have to prepare the
initial states of the qubits carefully. For a non-Gaussian field,
the uncertainty principle serves only a sufficient condition of
its entanglement.

\begin{figure} [ht]
{\bf (a)}\hskip5cm{\bf (b)}
\centerline{\psfig{figure=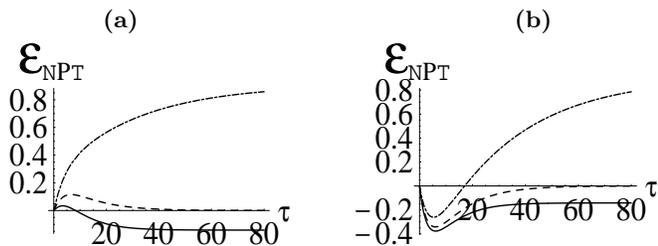,width=9.5cm,height=3.0cm}}
\caption{Dynamics of entanglement for the two qubits. The
entanglement  is plotted as a function of the dimensionless
interaction time $\tau=\gamma{t}$ for $n=2.4$ and three different
values of $c$: $c=1.58$ (solid curve), $c=c_{ss}(2.4)=1.804$
(dashed curve) and $c=\sqrt{n^2-1}=2.18$ (dot-dashed curve).  The
initial state is $\ket{gg}_{12}$ {\bf (a)} while it is
$\ket{ee}_{12}$ {\bf (b)} .} \label{benattipicture}
\end{figure}

Let us investigate the entanglement condition (\ref{benatti-c}).
In order to do it, we further assume a special case of $n=m$ and
$\gamma_{1}=\gamma_{2}=\gamma$ for a short while as it is not
straightforward to solve the general dynamic equation
(\ref{benatti}). In this case, Eq.(\ref{mycondition}) becomes
$n-1<c$.  For the density matrix elements
$\rho_{ijhk}=\sand{ij}{\rho_{12}}{hk}$ ($i,j,h,k=e,g$), we find
the coupled Bloch equations
%\begin{widetext}
\begin{equation}
\label{blochequations}
\begin{aligned}
&{\dot\rho}_{eeee}=\gamma\left[-2n^{1}_{1}\rho_{eeee}+n^{0}_{1}(\rho_{egeg}+\rho_{gege})+c\rho_{eegg}\right],\\
&{\dot\rho}_{egeg}=\gamma\left[n^{0}_{1}(1-\rho_{gege}) +\rho_{eeee}-n^{1}_{3}\rho_{egeg}-c\rho_{eegg}\right],\\
&{\dot\rho}_{gege}=\gamma\left[n^{0}_{1}(1-\rho_{egeg}) +\rho_{eeee}-n^{1}_{3}\rho_{gege}-c\rho_{eegg}\right],\\
&{\dot\rho}_{eegg}=-\gamma\left[n^{1}_{2}\rho_{eegg}-c(1/2
-\rho_{gege}-\rho_{egeg})\right],
\end{aligned}
\end{equation}
%\end{widetext}
where, $n^{l}_{k}=\frac{k}{2}(n-1)+l$ and, by hermiticity,
$\rho_{ggee}=\rho_{eegg}$. All the other matrix elements are
decoupled from these equations. The normalization condition
determines $\rho_{gggg}$. Solving Eqs.~(\ref{blochequations}), we
find the dynamics of the entanglement between the two qubits as
shown in Fig.~\ref{benattipicture}.

By inspection of Fig.~\ref{benattipicture} {\bf (a)}, it is
apparent that, even for $n-1<c$, the long time behavior of the
entanglement function can lead to a separable qubit state. This
is due to the fact that the condition (\ref{mycondition}) for the
dynamical entanglement-transfer does not give information about
the steady-state entanglement. The criterion (\ref{benatti-c}) for
entanglement due to the interaction with a Markovian environment
is, indeed, relative to the creation of quantum correlations in an
initially separable state.  The sufficient condition
(\ref{benatti-c}) to entangle two qubits comes from a positive
gradient of ${\cal E}_{NPT}$ at $t=0$. This is the case for
$\rho_{12}(0)=\ket{gg}_{12}\!\bra{gg}$  but not for
$\rho_{12}(0)=\ket{ee}_{12}\!\bra{ee}$, for example. For this
initial state, $\partial_{t}{\cal E}_{NPT}<0$ at $t=0$, as seen in
Fig.~\ref{benattipicture} {\bf (b)}.  In fact, if we start with
excited states for qubits the sufficient condition
(\ref{benatti-c}) leads to $c>n+1$, which is physically
meaningless as the CP condition and the uncertainty principle
impose $c^2\leq n^2-1$ (we remind that $n>{0}$). Does it impose
that qubits initially in their excited states will never be
entangled during their dynamics? The answer is `no'.  The
condition (\ref{benatti-c}) is only a sufficient condition and
Fig.~\ref{benattipicture} {\bf (b)} clearly shows that the qubits
can be entangled at a later period of the dynamical evolution. A
similar result is found for when one qubit is prepared
in $\ket{g}$ while the other is in $\ket{e}$.

It is, thus, interesting to investigate the conditions to entangle
qubits in their steady state. We now lift the temporary condition
$n=m$ and $\gamma_1=\gamma_2$ while keeping
$\modul{c_{1}}={\modul{c_{2}}}$ and find the boundary value
$c_{ss}(n,m)$ of the correlation parameter $c$ beyond which we
are sure that the qubit steady state is inseparable. To find
$c_{ss}(n,m)$ we have to look at the asymptotic behavior of the
entanglement function that can be found solving the
generalization of Eqs.~(\ref{blochequations}) to
$n\neq{m},\,\gamma_{1}\neq\gamma_{2}$ and looking for steady-state
solutions. Then, the condition
%\begin{equation}
%\label{wonmincondition}
%\left\{
%\begin{aligned}
$\lim_{t\rightarrow\infty}{\cal
E}_{NPT}(t,n,m,c)\vert_{c=c_{ss}(n,m)}=0$
%&\lim_{t\rightarrow\infty}\partial_{M}{\cal E}_{NPT}(t,c_{ss}(N))>0.
%\end{aligned}
%\right.
%\end{equation}
 fully characterizes the boundary value. We find
\begin{equation}
\label{boundary} c_{ss}(n,m)=-\frac{n\gamma_{1}+m\gamma_{2}}{
2\sqrt{\gamma_{1}\gamma_{2}}}\left[\frac{\mu(n,m)-\sqrt{\nu(n,m)}}{n^2m^2+(m-n)^2}\right]^{\frac{1}{2}}
\end{equation}
with $\mu(n,m)=(nm-1)^2+(nm+1)-(n-m)^2$  and
$\nu(n,m)=4nm+4(nm-1)^2-3(n-m)^2$. {\em The two qubits are
entangled at their steady state if and only if $c>c_{ss}(n,m)$.}

{\it Cavity quantum electrodynamic (CQED) system -} We consider a
CQED setup to illustrate the conditions for efficient entanglement
transfer. This model was recently suggested by Kraus and Cirac
~\cite{kraus} to show the possibility to entangle two identical
two-level atoms respectively trapped in two remote single-mode
cavities. The cavities are driven by a broadband two-mode
squeezed state. Here, we show that our general approach gives the
complete conditions to entangle the atoms. The cavity-driving
field coupling rates $\kappa$ are taken to be identical under the
identical cavity assumption. The adiabatic elimination described
above gives us the reduced atomic master equation. The
weak-coupling regime is now equivalent to the {\it bad-cavity
limit} in which the steady state of the cavity is the two-mode
squeezed state as in the case without atoms in the cavities.
Experimentally, $\kappa$ can not be taken large at will because
the relation $\Delta\omega_{ext}\gg\kappa$ represents a
constraint for the validity of our treatment. Typically, it is
$\Delta\omega_{ext}\simeq\kappa/6\simeq2\pi\times12\, MHz$ and
values as $(\Omega,\Gamma)/2\pi\simeq(20,\,3.5)\, MHz$ allow for
the validity of the bad-cavity regime and for the squeezed state
to build up inside the cavities~\cite{turchettekimble}. For the
CQED system here, the atoms are entangled in the steady state if
$c>c_{ss}=\frac{1}{n}\left(\sqrt{(n^2-1)^2+n^2}-1\right)$. This is
a severe constraint on the properties of the driving field. The
experimentally available source of squeezed light is, indeed,
quite bright, {\em i.e.}, $n$ large, so that $c_{ss}\rightarrow
\sqrt{n^2-1}$. This restricts the range of values of $c$ in which
the atomic {\em steady state} is entangled. Note that $c^2=n^2-1$
is satisfied by a pure state of the driving field.
%when the Gaussian field is a minimum-uncertainty state.

In the CQED model, the atoms may interact not only with the
single-mode cavity fields but also with other uncontrollable
reservoir through atomic spontaneous decay. Including the atomic
spontaneous emission of its rate $\Gamma$, the Louivillian remains
in the form (\ref{benatti}) but with $\gamma_{1,2}$, $n$ and $c$
replaced by $\gamma'_{1,2}=(2\Omega^2_{a1,b2}/ \kappa) (1+1/{\cal
C})$, $n'=m'=(n{\cal C}+1)/(1+{\cal C})$ and $c'=c \, {\cal C}
/(1+{\cal C})$ (${\cal C}=2\Omega^2/\Gamma\kappa$ is the {\it
cooperativity}~\cite{turchettekimble}). With these new parameters
and for the atoms initially in their ground states, the condition
to entangle the two {\em dynamic} atoms still remains as $(n-1)<c$
for the driving field. We find that {\em even with the atomic
spontaneous decay}, the atoms are guaranteed to be entangled for
periods of time by interaction with the entangled squeezed field
\footnote{The entanglement-transfer condition can
be generalized to the case of $n\neq m$ in inclusion of atomic
spontaneous decay. In this case, the cooperativity cancels out and
the entanglement condition is as in (\ref{mycondition}).
In fact, this condition is robust against not only the spontaneous
decay but also disparate couplings of the atoms with their cavity
modes.}.

\begin{figure} [ht]
%{\bf (a)}\hskip4cm{\bf (b)}
\centerline{\psfig{figure=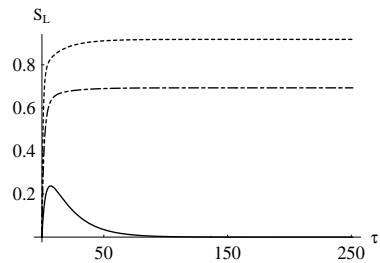,width=5.5cm,height=3.5cm}}
\caption{The linearized entropy $S_{L}$ is plotted against
$\tau=\gamma{t}$ for the pure environments $\sqrt{n^2-c^2}=1$ and
correlation $(n-c)^{-1}=2.5$ (solid line) and  for the mixed
environments $\sqrt{n^2-c^2}=1.4$ and $(n-c)^{-1}=2.5$ (dot-dashed
line) and $(n-c)^{-1}=3.5$ (dashed line). }
\label{benattipicture2}
\end{figure}

With these new parameters, it is always $n'^2>c'^2+1$, even for a
pure drive. Thus, inside the cavities, a pure two-mode squeezed
environment, where the discussion in~\cite{kraus} was centered, is
hard to be obtained in a realistic situation. It is thus worth
addressing the effect of the purity of the quantum channel in this
example. As a measure for the purity of the atoms, we take the
{\it linearized entropy}
$S_{L}=4/3\left(1-Tr_{12}\left\{\rho^2_{12}\right\}\right)$ that
ranges from 0 (pure states) to 1 (maximally mixed ones). Only the
interaction with a pure squeezed environment realizes a pure
atomic steady state. The dynamics of the linearized entropy $S_L$
for the atoms initially in their ground states, are shown in
Fig.~\ref{benattipicture2}. The higher is $\sqrt{n^2-c^2}$, the
more mixed the driving field and the higher is $(n-c)^{-1}$ and
the more the environments are entangled\cite{leekimmunro}. It is
apparent that only a slight departure from the pure state of the
driving field brings about the atoms extremely mixed as shown in
Fig.~\ref{benattipicture2}. However, as stated before, in a
realistic situation, $\Gamma\neq{0}$ so that it is not possible to
get the local environments in their pure correlated state. Hence,
the atomic steady state will always be mixed. When ${\cal C}\gg1$,
a nearly pure steady state may be obtained, however, the bad
cavity limit may not be used and the probability to feed the
cavities becomes small \cite{turchettekimble}.

{\it Remarks - }In this paper, we investigated the conditions to
entangle two remote qubits dynamically and in their steady state,
addressing the case of Markovian interaction with their local
environments, which are driven by a quantum channel.  We found
that the entanglement of the Gaussian environments is not only a
necessary but also a sufficient condition to see entanglement of
the qubits for some period of their evolution, provided the qubits
are appropriately prepared at the initial instance. We found the
boundary value of the correlation parameter of the quantum channel
only beyond which the steady state of the qubits is entangled.

%%%%%%%%%%%%%%%%%%RINGRAZIAMENTI E FINANZIAMENTO%%%%%%%%%%%%%%%%%%%%%%%%%%%%%%%

%\section*{Acknowledgments}
{\it Acknowledgments - }We thank Prof. S. Swain for fruitful
discussions. This work was supported by the UK Engineering and
Physical Science Research Council, Korean Research Foundation
(2003-070-C00024) and the International Research Centre for
Experimental Physics.
%%%%%%%%%%%%%%%%%%%%%%%%%%%%%%%%%BIBLIOGRAFIA%%%%%%%%%%%%%%%%%%%%%%%%%%%%%%%%%%%%%%%%%%%%%%%%%%%%%%%%%%%%%%%%%%%%%%%%%%%%%

\end{document}